\documentclass[10pt]{article}
\usepackage{geometry}                
\usepackage[parfill]{parskip}
\usepackage{amssymb}

\usepackage{amsmath}

\title{SQUEEZED STATES AND SYMPLECTIC TRANSFORMATIONS}
\author{C. V. Sukumar \\{\em Wadham College,}\\{\em University of Oxford, Oxford OX1 3PN, U.K. }
}

\begin{document}
\maketitle

\begin{abstract}
It is shown that the time evolution of the squeezed and displaced state may be obtained by solving the Heisenberg equation of motion of an appropriate operator and finding the eigenstates of the time evolved operator. The connection between symplectic transformations and squeezing is explored.

\bigskip

\centerline{\it PACS INDICES: 03.65, 42.50Dv}

\end{abstract}

\section{Heisenberg equation of motion}

It is well known that the squeezed and displaced states of the harmonic oscillator are eigenstates of a linear combination of the creation and annihilation operators of the oscillator with non-zero eigenvalue (Yuen 1975, Henry and Glotzer 1988). The properties of squeezed states have been discussed extensively,see for example: Ekert and Knight 1989, Gong and Aravind 1990, Gersch 1992. The squeezed state is a limiting case of the squeezed and displaced state corresponding to eigenvalue zero. Since the Heisenberg equations of motion for the position and momentum operators of a simple harmonic oscillator can be solved, it is possible to obtain the time evolution of the squeezed and displaced state by solving the Heisenberg equation of motion for a linear combination of position and momentum operators and then finding the eigenstates of the time evolved operator and establishing a relation between the two states. We outline such a procedure and also show that there is a connection between squeezing and symplectic transformations of position and momenta. For simplicity of notation, Planck's constant $\hbar$, the mass parameter $m$ and the frequency of the oscillator $\omega$ are all set equal to 1. In the first part of this paper we establish a relation between the eigenstates of a Heisenberg operator and the solutions of the time dependent Schr\"odinger equation. We then use this relation to obtain the time evolution of the squeezed and displaced state of the harmonic oscillator. In the final part of the paper we consider the connection between squeezing and symplectic transformations. 

Consider the eigenvalue equation for an operator A given by
\begin{equation}
A(t)\ \Phi_{j}(t)\ =\ \Lambda_{j}(t)\ \Phi_{j}(t).\label{}
\end{equation}
The Heisenberg equation of motion for an operator A which has no explicit time dependence is 
given by
\begin{equation}
i \ \frac{\partial A}{\partial t} \ = \ [A,H] \label{}
\end{equation}
where $H$ is a time independent Hamiltonian. Equation (2) has the formal solution
\begin{equation}
A(t)\ = \ {\rm e}^{iHt}\ A(0)\ {\rm e}^{-iHt}. \label{}
\end{equation}
The eigenvalue equation at time t=0 given by
\begin{equation}
A(0) \Phi_{j}(0)\ =\ \Lambda_{j} \Phi_{j}(0) \label{}
\end{equation}
may be rewritten using the solution given by equation (3) in the form
\begin{equation}
A(t)\ \bigl[ {\rm e}^{iHt}\ \Phi_{j}(0) \bigr]\ =\ \Lambda_{j}\ \bigl[{\rm e}^{iHt} \ \Phi_{j}(0) \bigr] \label{}
\end{equation}
which shows that the eigenvalues $\Lambda_{j}$ of $A(0)$ are also eigenvalues of $A(t)$ and the eigenstates of $A(t)$ are given by 
\begin{equation}
\Phi_{j}(t)\ \sim\ {\rm e}^{iHt} \ \Phi_{j}(0). \label{}
\end{equation}
In the above equation the $\sim$ symbol is used because a comparison of equations (1) and (5) shows that it is possible for the two normalized states to differ by a  phase factor which depends only on time but not any other variables that $H$ is a function of, as will be demonstrated explicitly in the next paragraph. Since the eigenvalues are time independent we can make the identification that $\Lambda_{j}(t)$ in equation (1) has no dependence on t, i.e. $\Lambda_{j}\equiv \Lambda_{j}(t)$. It may not always be convenient to solve equation (4) and then use equation (6) to find the eigenstates at time t as $\Phi_{j}(0)$ may be a complicated linear superposition of eigenstates of the Hamiltonian and the evaluation of the right hand side of equation (6) in closed form may be difficult. Under certain circumstances it may be possible to find $A(t)$ directly from equation (3) and solve the eigenvalue equation for $A(t)$ given by  equation (1) to find the eigenstates and eigenvalues of $A(t)$. The equation satisfied by $\Phi_{j}(t)$ may be found by differentiating equation (1) with respect to time and using equation (2) to give 
\begin{equation}\big[ A-\Lambda_{j} \big]\ \big[H\ \Phi_{j}(t)\ +\ i \frac{\partial \Phi_{j}(t)} {\partial t} \big]\ =\ 0. \label{}
\end{equation}
This equation is clearly satisfied if 
\begin{equation}
\big[H \ +\ i\ \frac{\partial} {\partial t} \big]\ \Phi_{j}(t)\ =\ \mu_{j}(t)\ \Phi_{j}(t) \label{}
\end{equation}
in which $\mu_{j}(t)$ is an arbitrary function of time. The solutions of equation (8) may be given in the form
\begin{equation}
\Phi_{j}(t)\ =\ {{\rm e}^{-i\int^{t}_{0} \mu_{j}(t') { dt'}}} \ F_{j}(t) \label{}
\end{equation}
where
\begin{equation}
\big[ H \ + \ i \ \frac{\partial }{\partial t} \big]\ F_{j}(t)\ =\ 0. \label{}
\end{equation}
It is now possible to identify that
\begin{equation}
\Psi_{j}(t)\ =\ F_{j}(-t) \label{} 
\end{equation}
is the Schr\"odinger state evolving under the influence of $H$ from the state $\Phi_{j}(0)=F_{j}(0)=\Psi_{j}(0)$. Thus we can relate the Schr\"odinger state $\Psi_{j}(t)$ to the solutions of equations (1) and (3) in the form
\begin{equation}
\Psi_{j}(t)\ =\ {\rm e}^{-iHt}\ \Psi_{j}(0)\ ={\rm e}^{i \delta (t)}\ \Phi_{j}(-t) \label{}
\end{equation}
where
\begin{equation}
 \delta (t)\ =\ \int^{-t}_{0} \mu_{j}(t')\  dt' .\label{}
\end{equation}
We have shown that apart from a time dependent phase factor, the solution $\Psi_{j}(t)$ of the Schr\"odinger equation for $H$ may be constructed from the solution $\Phi_{j}(-t)$ of the eigenvalue equation for the time dependent Heisenberg operator $A(-t)$ by propagating backwards in time. To find the phase factor $\delta(t)$ we can expand both $\Psi_{j}(t)$ and $\Phi_{j}(t)$ in terms of the eigenstates of $H$ with eigenvalues $E_{n}$ which are denoted by $|n\rangle$, $n=0,1,2,...$. Let
\begin{align}
\Psi_{j}(t)\ &=\ \sum_{n} a_{n}(t) |n\rangle, \notag \\
\Phi_{j}(t)\ &=\ \sum_{n} b_{n}(t) |n\rangle.\label{}
\end{align}
It is clear from equations (12) and (13) that
\begin{align}
a_{n}(t)\ &=\ b_{n}(-t)\ {\rm e}^{i\ \delta(t)},\notag \\
a_{n}(0)\ &=\ b_{n}(0) \label{}
\end{align}
In terms of the eigenvalues $E_{n}$ of $H$ 
\begin{equation}
a_{n}(t)\ =\ a_{n}(0) \ {\rm e}^{-iE_{n}t}. \label{}
\end{equation}
Hence
\begin{equation}
{\rm e}^{i\  \delta(t)}\ = \ {\rm e}^{-iE_{n}t} \frac{b_{n}(0)}{ b_{n}(-t)}.\label{}
\end{equation}
The simplest way to find $\delta(t)$ is to choose $n=0$ so that
\begin{equation}
{\rm e}^{i\ \delta(t)}\ =\ {\rm e}^{-iE_{0}t}\ \ \frac{\langle0|\ \Phi_{j}(0)\rangle}{ \langle 0|\Phi_{j}(-t)\rangle} .\label{}
\end{equation}
Thus the Schr\"odinger state $\Psi_{j}(t)$ may be obtained from the solution $\Phi_{j}(t)$ of equations (1) and (3) using the relation
\begin{equation}
\Psi_{j}(t)\ =\ {\rm e}^{-iE_{0}t}\  \frac{\langle 0|\ \Phi_{j}(0)\rangle}{\langle 0|\Phi_{j}(-t) \rangle}\ \ \Phi_{j}(-t) .\label{}
\end{equation}

\section{Time evolution of squeezed states of simple harmonic oscillator}

We now illustrate the procedure by studying the time evolution of the squeezed and displaced state of the simple harmonic oscillator. Consider the harmonic oscillator governed by the Hamiltonian
\begin{equation}
H\ =\ {\frac{1}{2}} \bigl( P^{2} \ +\ X^{2} \bigr).\label{}
\end{equation}
The commutation relation $[X,P]=i$ maybe used to establish the Heisenberg equations of motion for the operators $X(t)$ and $P(t)$ which may then be solved to give
\begin{align}
X(t)\ &=\ X(0)\ \cos t\ +\ P(0)\ \sin t,\notag \\
P(t)\ &=\ P(0)\ \cos t\ -\ X(0)\ \sin t .\label{}
\end{align}
Let
\begin{equation}
A(t)\ =\ i\alpha P(t)\ +\ \beta X(t). \label{}
\end{equation}
Then using the operator solutions given by equation (21) and adopting the notation that $X(0)\equiv X, P(0)\equiv P,\alpha(0)\equiv\alpha, \beta(0)\equiv\beta,$ $A(t)$ may be given in the form
\begin{equation}
A(t)\ =\ i\alpha (t) P\ +\ \beta (t) X, \label{}
\end{equation}
where
\begin{align}
\alpha (t)\ &=\ \alpha \cos t\ -\ i \beta \sin t,\notag \\
\beta(t)\ &=\ \beta \cos t\ -\ i \alpha \sin t.\label{}
\end{align}
Using $P=i\frac{\partial}{\partial X}$ it is easy to verify that the eigenstates of $A(t)$ with eigenvalue $\lambda$ are given by
\begin{equation}
\Phi(t)\ \sim{\rm exp}\bigg[- \frac{\beta (t)}{ 2\alpha(t)}\Big(X-\frac{\lambda}{\beta (t)}\Big)^{2}\bigg]. \label{}
\end{equation}
It is clear that $\Phi(0)$ represents a squeezed and displaced state with squeeze parameter $S$ and displacement parameter $D$ given by
\begin{equation}
S(0)\ =\ \frac{\beta}{ \alpha}\ ,\ D(0)\ =\ \frac{\lambda}{\beta}\ .\label{}
\end{equation}
The time evolution of the squeezed and displaced state may now be obtained using equations (19),(25) and (24). It is easily seen that the squeeze and displacement parameters of the time evolved state are given by
\begin{align}
 S(t)\ &=\ \frac{\beta(-t)}{\alpha (-t)}\ =\ \frac{\bigl( \beta \cos t \ +\ i\alpha \sin t\bigr)}{\bigl( \alpha \cos t\ +\ i\beta \sin t\bigr)},\notag \\
 D(t) &=\ \frac{\lambda}{\beta(-t)} \ =\ \frac{\lambda}{\bigl( \beta \cos t\ +\ i\alpha \sin t\bigr)}	. \label{}
\end{align}
These equations may be expressed in terms of the squeeze and displacement parameters at time $t=0$ in the form 
\begin{align}
S(t)\ &=\ \frac{\bigl(S(0)\cos t\ +\ i\sin t\bigr)}{\bigl(\cos t\ +\ iS(0)\sin t\bigr)},\notag \\
D(t)\ &=\ \frac{D(0)S(0}{\bigl(S(0)\cos t\ +\ i\sin t\bigr)} .\label{}
\end{align}
The squeezed and displaced states are displaced gaussians with complex exponents. The normalization factor associated with these states and the overlap integrals appearing in equation (19) depend upon $t$ and may be readily evaluated. The normalized squeezed and displaced state may finally be given in the form
\begin{align}
\Psi_{sd}(x,t)\ &=\ N(t)\  {\rm exp}\bigl[-\ \frac{1}{2} S(t)\bigl(X^{2}\ -\ 2D(t)X\ +\ D(t)D(0)\cos t\bigr)\bigr], \notag \\
N(t)\ &=\ \bigl( \frac{S(0)}{\pi}\bigr)^{\frac{1}{4}}\ \bigl( \cos t\ +\ iS(0)\sin t\bigr)^{-\frac{1}{2}}\ .\label{}
\end{align}
The squeezed vacuum state corresponds to choosing $D(0)=0$ and coherent states correspond to choosing $S(0)=1$ in the general expression given above. It may be shown that the probability distribution associated with $\Psi_{sd}(t)$ is given by
\begin{align}
\Psi_{sd}^{\star} \Psi_{sd}\ &=\ \bigl( \frac{\gamma}{\pi}\bigr)^{\frac{1}{2}}\ {\rm exp}\bigl[ -\gamma\bigl( X\ -\ D(0)\cos t\bigr)^{2}\bigr] , \notag \\
\gamma\ &=\ \frac{S(0)}{\bigl(\cos^{2}t\ + S^{2}(0)\ \sin^{2}t\bigr)}\ .\label{}
\end{align}
This probability distribution corresponds to a wavepacket whose centre oscillates with classical frequency 1 between the classical turning points $\pm D(0)$ and the shape of the wavepacket undergoes a time dependent squeeze. The simple feature that $\Psi_{sd}$ is a gaussian with time dependent squeeze and displacement parameters suggests that there may be a connection between squeezing and canonical transformations in classical mechanics. Towards this end we consider a linear transformation from a set of canonical coordinates $X_{1}$ and $P_{1}$ satisfying the commutation relation $\bigl[X_{1},P_{1}\bigr]=i$ to a new set of coordinates $X_{2}$ and $P_{2}$ through a mapping of the form
\begin{align}
X_{2}\ &=\ aX_{1}\ +\ bP_{1}\ ,\notag \\
P_{2}\ &=\ cX_{1}\ +\ dP_{1} .\label{}
\end{align}
It is clear that if 
\begin{equation}
ad-bc\ =\ 1\ , \label{}
\end{equation}
then $\bigl[X_{2},P_{2}\bigr]=i$ and the new coordinates also satisfy the same commutation relation as the old coordinates. Transformations of the form of equation (31) with unit determinant are symplectic transformations (Goldstein 1980). It is well known that apart from the unit matrix there are only three other matrices of dimension 2 which have unit determinant (Schiff 1968) and they may be written in terms of the Pauli spin matrices
\begin{equation}
\sigma_{x}\ =\ \left(\begin{array}{cc}0&1\\ 1&0\end{array}\right),\qquad
\sigma_{y}\ =\ \left(\begin{array}{cc}\ 0& -i\\ +i&\ 0\end{array}\right),\qquad
\sigma_{z}\ =\ \left(\begin{array}{cc}1&\ 0\\ 0&-1\end{array}\right) \label{}
\end{equation}
in the form
\begin{equation}
{\rm e}^{-\theta \sigma_{x}}\ =\ \left(\begin{array}{cc}\cosh\theta & \sinh \theta \\ \sinh \theta & \cosh \theta \end{array}\right),\ {\rm e}^{i\nu \sigma_{y}}\ =\ \left(\begin{array}{cc}\ \cos \nu & \sin \nu \\ -\sin \nu &\cos \nu\end{array}\right),\ {\rm e}^{\rho \sigma_{z}}\ =\ \left(\begin{array}{cc}{\rm e}^{\rho} &0\\ 0 & {\rm e}^{-\rho}\end{array}\right).\label{}
\end{equation}
Any matrix in two dimensions with unit determinant may thus be constructed from the exponentials of the matrices $1,\sigma_{x},\sigma_{y}$ and $\sigma_{z}$. If we now consider the eigenvalue equation 
\begin{equation}
\bigl(i\alpha P_{1}\ +\ \beta X_{1}\bigr) \Psi \ =\ 0 \label{}
\end{equation}
then
\begin{equation}
\Psi\ =\ {\rm exp}\bigl(-\frac{1}{2} S_{1} X_{1}^{2}\bigr)\  ,\  S_{1}\ =\ \frac{\beta} {\alpha}  \label{}
\end{equation}
is a squeezed state with squeeze parameter $S_{1}$. In terms of $X_{2}$ and $P_{2}$ equation (35) may be written in the form
\begin{equation}
\Bigl[i\bigl(\alpha a\ +\ i\beta b\bigr)P_{2}\ +\ \bigl(\beta d\ -\ i\alpha c\bigr) X_{2}\Bigr]\ \Psi\ =\ 0 \label{}
\end{equation}
with the solution
\begin{equation}
\Psi\ =\ {\rm exp}\bigl(-\frac{1}{2} S_{2} X_{2}^{2}\bigr)\  ,\ \ S_{2}\ =\ \frac{\beta d\ -\ i\alpha c}{ \alpha a\ +\ i\beta b}\ , \label{}
\end{equation} 
which is also a squeezed state with squeeze parameter $S_{2}$. Thus we can see that when two sets of coordinates are connected by a symplectic transformation then a squeezed state with respect to the old coordinate transforms into a squeezed state with respect to the new coordinate with a simple transformation between the squeeze parameters given by
\begin{equation}
S_{2}\ =\ \frac{S_{1}d\ -\ ic}{ a\ +\ iS_{1}b}. \label{}
\end{equation}
Comparison of equations (21) and (31) shows that 
\begin{equation}
a\ =\ \cos t,\ b\ =\ \sin t,\ c\ =\ -\sin t,\   d\ =\ \cos t.\label{}
\end{equation}
Since $(ad-bc)=1$ it is clear that the time evolution of the position and momentum operators given by equation (21) is a symplectic transformation. The time evolution of the squeeze parameter given by equation (28) is of the form given by equation (39). Comparison with equation (34) shows that time evolution is a symplectic transformation of the type induced by $\sigma_{y}$. In this paper we have outlined a novel procedure for studying the time evolution of the Schr\"odinger states. We have shown that the time evolution of the squeezed and displaced state may be studied using this method. We have shown that under a symplectic transformation a squeezed state transforms into a new squeezed state with a transformed squeeze parameter. For a 1-dimensional system there are three sets of symplectic transformations possible apart from the identity transformation. We have shown that the time evolution of the squeezed state may be related to one of the three non-trivial symplectic transformations.

\bigskip

\section{Acknowledgment}

I thank David Brink for many interesting discussions on symplectic transformations.

\bigskip

\section{REFERENCES}

1. H.P.~Yuen, Phys. Rev. A {\bf 13}, 2226-2243 (1975).

2. R.W.~Henry and S.C.~Glotzer, Am. J. Phys. {\bf 56} 318-328 (1988).

3. A.K.~Ekert and P.L.~Knight, AM. J. Phys. {\bf 57} 692-697 (1989).

4. J.J.~Gong and P.K.~Aravind, Am. J. Phys. {\bf 58} 1003- 1006  (1990).
 
5. H.A.~Gersch, Am.J.Phys. {\bf 60} 1024-1030 (1992).

6. H.~Goldstein, Classical Mechanics ({\it Addison-Wesley},Singapore) 393-396 (1980).

7. L.I.~Schiff, Quantum Mechanics ({\it McGraw-Hill}, New York) 206-207(1968).

\end{document}